\documentclass[letterpaper, 10 pt, conference]{ieeeconf}  

\IEEEoverridecommandlockouts                              
\usepackage{amsmath,amsfonts,amssymb,amsthm,psfrag,enumerate}
\usepackage{graphicx}

\usepackage{algorithm}
\usepackage{framed}
\usepackage{algpseudocode}

\newcommand{\bb}{\ensuremath\begin{bmatrix}}
\newcommand{\eb}{\ensuremath\end{bmatrix}}
\newcommand{\R}{\mathbb{R}}

\newcommand{\N}{\mathbb{N}}

\def\defeq{\mathrel{\mathop:}=}

\usepackage[framemethod=tikz]{mdframed}
\newtheorem{thm}{Theorem}

\newtheorem{prp}{Proposition}
\definecolor{mygray}{RGB}{248,248,250}
\newtheoremstyle{mystyle}
  {\topsep}
  {\topsep}
  {}
  {}
  {\bfseries}
  {.}
  {5pt plus 1pt minus 1pt}
  {{\color{black}\thmname{#1}~\thmnumber{#2}}\thmnote{\,--\,#3}}%
\theoremstyle{mystyle}
\newmdtheoremenv[%
  backgroundcolor=mygray,%
  linecolor=black,%
  leftmargin=0pt,%
  innerleftmargin=5pt,%
  innerrightmargin=5pt,%
  ]{probx}{Problem}

\title{\LARGE \bf
On the Design of an Intelligent Speed Advisory System for Cyclists
}

\author{Yingqi Gu$^{1}$, Mingming Liu$^{1,\star}$, Matheus Souza$^{2}$ and Robert N. Shorten$^{1}$
\thanks{$^{1}$Y. Gu, M. Liu and R. N. Shorten are with the School of Electrical and Electronic Engineering, University College Dublin, Ireland.}%
\thanks{$^{2}$M. Souza is with the School of Electrical and Computer Engineering, University of Campinas, Brazil.}%
\thanks{$^{\star}$Corresponding author. Email: {\tt mingming.liu@ucd.ie}}%
}

\begin{document}

\maketitle
\thispagestyle{empty}
\pagestyle{empty}


\begin{abstract}
	
Traffic-related pollution is becoming a major societal problem globally. Cyclists are particularly exposed to this form of pollution due to their proximity to vehicles' tailpipes. In a number of recent studies, it is been shown that exposure to this form of pollution eventually outweighs the cardio-vascular benefits associated with cycling. Hence during cycling there are conflicting effects that affect the cyclist. On the one hand, cycling effort gives rise to health benefits, whereas exposure to pollution clearly does not. Mathematically speaking, these conflicting effects give rise to convex utility functions that describe the health threats accrued to cyclists. More particularly, and roughly speaking, for a given level of background pollution, there is an optimal length of journey time that minimises the health risks to a cyclist. In this paper, we consider a group of cyclists that share a common route. This may be recreational cyclists, or cyclists that travel together from an origin to destination. Given this context, we ask the following question. What is the common speed at which the cyclists should travel, so that the overall health risks can be minimised? We formulate this as an optimisation problem with consensus constraints. More specifically, we design an intelligent speed advisory system that recommends a common speed to a group of cyclists taking into account different levels of fitness of the cycling group, or different levels of electric assist in the case that some or all cyclists use e-bikes (electric bikes). To do this, we extend a recently derived consensus result to the case of quasi-convex utility functions. Simulation studies in different scenarios demonstrate the efficacy of our proposed system.	
	
\end{abstract}

\section{Introduction}

In recent years, traffic-induced air pollution issues have been recognised  as one of the major threats for human health in cities \cite{laumbach2012respiratory,anderson2012clearing, tainio, krzyzanowski2005health}. Air contaminants, such as CO, $\textrm{NO}_\textrm{x}$, and particulate matter (PM), emitted from tailpipes of conventional vehicles (i.e. vehicles using internal combustion engines for propulsion), can result in serious health concerns for the general public. For instance, research in \cite{anderson2012clearing} shows that the mortality rate for people living in the most polluted cities can be 29\% more than those living in the least polluted cities based on data in the past several decades. A recent work in \cite{chen2017living} also indicates that living near major roads might adversely affect cognition, leading higher incidence to suffer neurodegenerative diseases (e.g. dementia, Parkinson). In reality, cyclists are more vulnerable to this form of pollution as they are usually closer to the tailpipes than other road users, and due to their elevated breathing rate \cite{bigazzi2014review}. Several papers have recently appeared that have begun to address these problems. Roughly speaking, these papers either modify cars' behaviour when close to pedestrians or cyclists, or suggest strategies that enable cyclists or pedestrians to protect themselves from the effects of pollution. For example, in \cite{sweeney2017cyberphysics} e-bike electrical assist is used to regulate the breathing rate of the cyclists. The interest reader is referred to the following for related work \cite{liu16, gu2018pedestrian, gu2016smart, schlote2013cooperative, naoum2017smart, sweeney2017cyberphysics,herrmann2018new} for further information on this topic. \newline

Our starting point in this paper is the work described in \cite{sweeney2017cyberphysics}. As mentioned, the authors in this paper attempt to regulate the breathing rate of a cyclist (or an e-bike) by modulating the amount of electrical assist provided to the cyclist. The rough idea is to bring down this rate in areas of elevated pollution. While this idea makes sense, breathing rate is only part of the story when discussing the health benefits of cycling. Clearly, the rate and duration of inhalation of ``dirty'' air is bad for the health of the cyclist. On the other hand, the cardio-vascular benefit of cycling is proportional to the amount of cycling effort. Together, these complementary effects determine the health benefits of cycling for an individual, and recently they were the subject of a study presented in \cite{tainio}, the high-level results of which are depicted in Figure 1. Clearly, these complementary effects give rise to a convex relationship characterising the health benefit of cycling. Using the nomenclature of \cite{tainio}, beyond the {\em breakeven point}, cycling is harmful to health, with maximum benefit realised at the so called {\em tipping} point. Clearly, journeys of tipping point duration are most beneficial to the cyclist.\newline

\begin{figure}[htb]
	\begin{center}
		\includegraphics[width=3.4in, height = 2.4in]{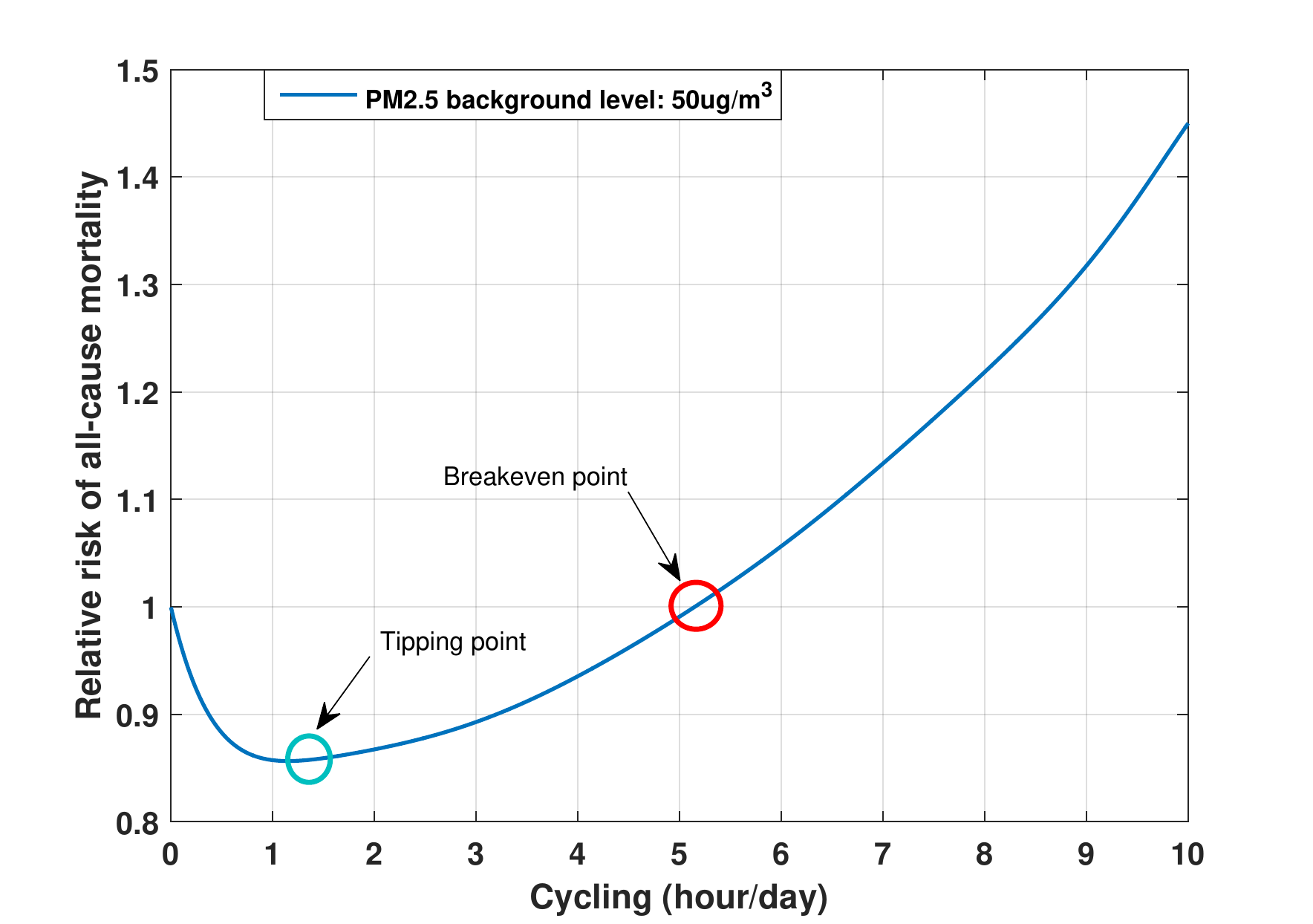}
		\caption{Recreated from Figure 1 in \cite{tainio}. This figure shows the tipping point and break-even point as measured by the relative risk for all-cause mortality combining the effects of air pollution (at $50 {\rm \mu g/m^3}$ ${\rm PM_{2.5}}$) and physical activity (cycling).}\label{risk}
	\end{center}
\end{figure}

\noindent {\bf Comment:} The precise nature of the curve depicted in Figure 1, hitherto referred to as a {\em utility function}, is an approximation that depends on a number of factors. Apart from average speed, background pollution levels, the health and fitness of the cyclist plays an important role. Furthermore, for e-bikes, the addition of electrical assist has the effect of {\em stretching} the curve; that is, for a given subject and route, the amount of effort is reduced as assist is provided, but also the rate at which pollution is inhaled \cite{sweeney2017cyberphysics} due to reduced breathing rate. Indeed - one may make use of this stretching to ensure that the cyclist is operating at the {\em tipping} point for a given journey. Notwithstanding this fact, the qualitative nature of the curve is correct - the longer one cycles, the greater the cardio-vascular benefit, but also the greater the amount of pollution inhaled. Clearly, knowledge of this curve for an individual opens up new possibilities to minimise the health threats of cycling for both individuals, and groups of cyclists.\newline

Given this basic setting we shall explore the following problem. We consider a group of cyclists that share a common route. This may be recreational cyclists, or cyclists that travel together from an origin to destination. We assume that each cyclist is characterised by a known utility function. Given this context, we ask the following question: what is the common speed at which the cyclists should travel, so that the overall health risks can be minimised? We formulate this as an optimisation problem with consensus constraints. More specifically, we design an intelligent speed advisory system that recommends a common speed to a group of cyclists taking into account different level of fitness of the cycling group, or different levels of electric assist in the case that some or all cyclists use electric bikes. Simulation studies in different scenarios demonstrate the efficacy of our proposed system. \newline

\noindent {\bf Contribution:} The design of speed advisory systems (SAS) for vehicles, and more advanced platooning systems, has a rich history in the automotive domain. We believe the system suggested here is the first of its type for cycling. In particular, as cycling, and e-bikes, become more popular, we believe that systems of this type may play an important role in a smart city context. To do this, we extend a recently derived consensus result to the case of quasi-convex utility functions \footnote{The function $f$ is said to be quasi-convex if, for every real $c$, $\left\lbrace x: x \in \Omega, f(x) < c \right\rbrace$ is convex, where $\Omega$ is a convex subset of $\R^n$ \cite{luenberger1968quasi}.}. This latter mathematical contribution may find wider used for consensus problems in an intelligent transportation system (ITS) context. \newline

The rest of the paper is organised as follows. Related works in literature are reviewed in Section \ref{RW}. System model and algorithm are presented in Section \ref{Math}. Simulation studies in different scenarios are discussed in Section \ref{Simulation}. Finally, we conclude the paper and in Section \ref{Conclusion}.

\section{Related Work} \label{RW}

In \cite{liu16}, a distributed SAS has been proposed to recommend a common speed for different types of vehicles to optimise their performance in ITS; namely, group emissions, or group battery consumptions, are minimised for conventional and electric vehicles, respectively. In there, cost functions were modelled using strictly convex functions for both emission generation and energy consumption to different types of vehicles. Optimisation problems were formulated that seek to minimise the costs with consensus constraints on speeds. To this end, an optimal distributed consensus algorithm was applied for all users in a manner that preserves privacy. Similar idea was then extended in \cite{griggs2018leader}, where two distributed SASs have been introduced with a target to recommending a common speed for a set of moving vehicles. In particular, the system was implemented using consensus based algorithm in a parallel networks that allows a way to obfuscate the input signal received by each vehicle via some noise. Rigorous proof was also provided to illustrate sufficient conditions on convergence of states in such a stochastic network. \newline

Concerning cyclists, the authors in \cite{herrmann2018new} have developed an optimisation algorithm for a set of plug-in hybrid electric vehicles (PHEVs) to dynamically mitigate the level of emissions around cyclists in virtual geographical boundaries (geofences). This was achieved by considering PHEVs as power-split devices, and the group of PHEVs in geofences are coordinated to automatically switch on/off their electric motors such that the overall emission around a cyclist can be maintained to a pre-defined safety level. The problem was formulated for each cyclist as an online optimisation problem with an emission budget constraint on PHEVs taking account of the background pollutant level and the likelihood of a cyclist's routing paths. Compared to \cite{liu16}, this work explores the actuation possibilities of PHEVs (i.e. by switching on/off electric motors) to maximise the environmental benefits for a single cyclist. Similar idea has been further explored in \cite{sweeney2017cyberphysics} where now a cyclist has the ability to access an e-bike with electrical assist if needed. An optimisation problem was formulated with an objective to provide better heart protection for cyclists. Roughly speaking, this was achieved by distributing more electric energy on e-bikes when cyclists entering areas with higher background pollutant levels, and in such a way that the ventilation rate of a cyclist can be indirectly controlled for better health benefits. \newline

%

\section{System Model and Algorithm}\label{Math}

\subsection{Model Assumptions}

\begin{figure}[htbp]
	\begin{center}
		\includegraphics[width=3.4in, height = 2.4in]{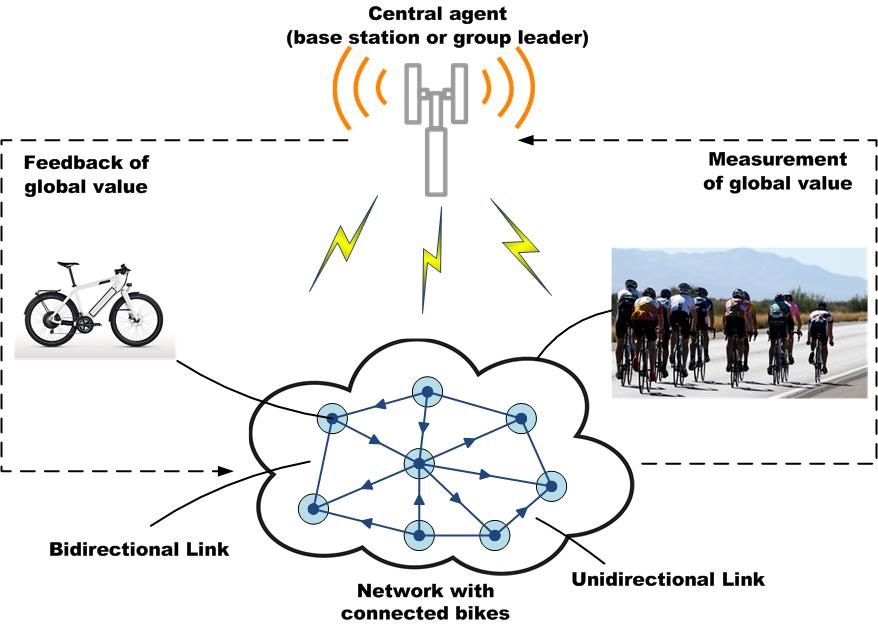}
		\caption{A schematic diagram of the proposed speed advisory system for a group of cyclists.}\label{schematic}
	\end{center}
\end{figure}

Our main goal in this paper is to devise a {\em speed-advisory system} that finds a common recommended speed for a fleet of $N$ bikes, including e-bikes, that share a common route. To do this, let us assume that each bike is equipped with a dedicated communication device, which is capable of receiving/transmitting messages between nearby bikes (e.g. using WLAN or Bluetooth), and road infrastructure through available communication channels (e.g. 3G/LTE networks) if applicable. In this context, each bike can send a broadcast signal to its neighbours (i.e. nearby bikes), and a limited amount of information to either infrastructure (e.g. a base station) if available, or the leader of the cycling group, who is acting as a central agent, when the road infrastructure is not available. After collecting of all information, the central agent will send back a broadcast signal to the entire network of bikes as a response. A schematic diagram of the proposed architecture is depicted in Figure \ref{schematic}. \\

\noindent \textbf{Comment: } The topology of the constructed communication network is inherently time-varying, especially considering different levels of transmission range, and the uncertainty of communication delays and failures among cyclists. We model the behaviour of information exchange in such a network using both unidirectional and bidirectional links as shown in Figure \ref{schematic}, where the unidirectional link represents the reachability of data in a specific direction. While it is not our primary focus to model the uncertainty of networks in this paper, we do require some specific properties on such stochastic networks as a prerequisite for the design of our algorithm. The details of which will be presented in the following sections. \newline

For simplicity, we shall require that all devices with the bikes can access to a common clock signal (e.g. GPS clock). Let $k \in \left\lbrace 1,2,3,... \right\rbrace$ be a discrete-time instant in which new information from bikes is collected and new speed recommendations are made. Denote by $N_k^i$ the set of nearby bikes (neighbours) of bike $i$. Let $s_i(k)$ denote the recommended speed of the bike $i \in \left\lbrace 1,2,\ldots,N \right\rbrace$ at time $k$, and $\textbf{s}(k)^{\textrm{T}} := \left[s_1(k), s_2(k), \ldots, s_{\textrm{N}}(k)\right]$ be the vector of recommended speeds of all bikes at time $k$, where the superscript T represents the transposition of the vector. In addition, we also assume that each cyclist $i$ is associated with a risk function $f_i$, which depends on a cyclist's daily travel times $t_i$. Each function $f_i$ is assumed to be strictly convex and has a global minimum point for optimal travel time $t_i^\star \in \left(0, t_i\right)$. Note that this assumption is in accordance to the shape of the curve depicted in Figure \ref{risk}.\\

\noindent \textbf{Comment: } Note that $f_i$'s are risk functions which contain sensitivity information associated with individual cyclists. For example, these functions could be used to discern fitness levels of individuals and perhaps some other health related information. For this reason, it is not desirable to share the $f_i$'s between cyclists, but rather to only allow a trusted node access to the $f_i$ information, e.g., a single leader or an ITS base station. This consideration is reflected in the architecture depicted in Figure \ref{schematic}.

\subsection{Problem Statement}

In this set-up, we wish to iteratively regulate the recommended speeds $\textbf{s}(k)$ to consensus (i.e. each component of the vector $\textbf{s}(k)$ is equalised) while minimising the overall risk of all-cause mortality, which combines the benefits of cycling with the negative effects coming from air pollution, from all cyclists. In practice, this can be achieved by assisting each cyclist with their bike's electric motor, and thus keeping their travelling speed as close as possible to the recommended common speed. In this context, we consider the main problem to be solved in this paper as follows.

\begin{probx} \label{prob}
Design a distributed SAS system for a network of bikes connected via a dedicated communication system, in order to recommend a common speed that minimises the overall risk of all-cause mortality due to exposure to air pollution of the whole group of cyclists.
\end{probx}

We note that in the above model each risk function $f_i$ is defined as a function of cycling time as in Figure \ref{risk}. However, in order to achieve consensus on speeds, we still need to factor each cyclist's travel distance $d_i$ in our problem formulation. Assuming that the travel distances $d_i$ of each cyclist is known {\em a priori} \footnote{An estimate could suffice in practical implementations.}, we can now formulate the main optimisation problem to be solved as
\begin{equation}\label{eq201}
\begin{array}{r}
\displaystyle \min_{t_1,\cdots,t_N \in \R_+} \sum_{i = 1}^N f_i(t_i) \vspace{0.2cm}  \\
\displaystyle {\rm s.t.} \quad \frac{t_i}{d_i} = \frac{t_j}{d_j}, \; i > j.
\end{array}
\end{equation}

\noindent \textbf{Comment:} The optimisation problem above is convex with the cycling times $t_i$ as decision variables. However, this problem has not been formulated as a consensus problem yet, which is difficult to be solved using our optimal consensus distributed algorithm to be discussed later. To solve this issue, we now define the new risk function $g_i(s_i) := f_i(d_i/s_i)$, for each $i$, which depends on speed $s_i$, and then we can reformulate the optimisation problem in \eqref{eq201} as follows
\begin{equation}\label{eq202}
\begin{array}{r}
\displaystyle \min_{s_1,\cdots,s_N \in \R_+} \sum_{i = 1}^N g_i(s_i) \vspace{0.2cm}  \\
\displaystyle {\rm s.t.} \quad s_i = s_j, \; i > j.
\end{array}
\end{equation}

\noindent \textbf{Comment:} Note that the optimisation problem \eqref{eq202} is now in consensus form after changing of functions. Although this problem formulation looks similar to our previous work in \cite{liu16}, as we shall see, each utility function $g_i$ is now strictly quasi-convex (i.e. not strictly convex). Nevertheless, this new mathematical formulation can still be solved in a distributed algorithm originally applied for strictly convex functions.

\subsection{Mathematical Results}

In this section, we present some auxiliary results that are used in this paper. The first result presents an important property for a particular optimisation problem.

\begin{prp} \label{prp01}
	Let $f \, : \, \R \to \R$ be a strictly convex ${\cal C}^2$ function that presents a (unique) global minimiser $y^\star \in \R$. For some open interval $\mathbb{I}$, let $h \, : \, \mathbb{I} \to \R$ be a strictly monotonic ${\cal C}^2$ function such that $y^\star \in h(\mathbb{I})$. Then, $g = f \circ h$ is a strictly quasi-convex function whose unique global minimiser is $x^\star = h^{-1}(y^\star)$.
\end{prp}

\begin{proof}
	Let us first show that $g$ is strictly quasi-convex. Take two arbitrary numbers $u,v \in \mathbb{I}$ such that $u < v$. From the strict convexity of $f$, it follows that
	\begin{align}\label{eq101}
	g(x) &= f\left(h(x)\right)\nonumber\\
	&< \max\Big\{ f\left(h(u)\right),f\left(h(v)\right)\Big\} =  \max\Big\{ g(u),g(v)\Big\}
	\end{align}
	holds for any $x \in (u,v)$, since $h(x)$ is either in $\left(h(u),h(v)\right)$, if $h$ is strictly increasing, or in $\left(h(v),h(u)\right)$, if $h$ is strictly decreasing. Hence, it follows that $g$ is strictly quasi-convex.
	
	Now, to prove the existence of a unique global minimiser for $g$, we first note that
	\begin{equation}\label{eq102}
	g'(x) = h'(x)f'(h(x)) = 0 \iff h(x) = y^\star,
	\end{equation}
	since $y^\star$ is the unique minimiser of $f$. This implies that $x^\star \defeq h^{-1}(y^\star)$ is the only stationary point for $g$. Additionally, as
	\begin{equation}\label{eq103}
	g''(x) = \big[h'(x)\big]^2 f''(h(x)) + h''(x)f'(h(x)), \; x \in \mathbb{I},
	\end{equation}
	we have that $g''(x^\star) > 0$, since $f$ is strictly convex. Thus, from the sufficient second order conditions \cite{Bazaraa_pnl}, it follows that $x^\star$ is a strict local minimiser of $g$. As $g$ is strictly quasi-convex, this implies that $x^\star$ is its unique global minimiser \cite{Bazaraa_pnl}. The proof is complete.
\end{proof}

We note that, from Proposition \ref{prp01}, it follows that each risk function $g_i$ is strictly quasi-convex and, moreover, each of them has a unique minimum; that is, each optimal trave time $t_i^\star$ is uniquely mapped onto an optimal associated speed $s_i^\star = d_i/t_i^\star$. Nevertheless, these observations alone are not sufficient to guarantee the existence of a unique global minimum for \eqref{eq202}, as the sum of strictly quasi-convex functions is not necessarily strictly quasi-convex. We further exploit the properties of \eqref{eq202} in the following proposition to show a condition on existence and uniqueness of solution to \eqref{eq202}.

\begin{prp} \label{prp02}
The speed $s^\star \in \R_+$ is the unique global minimiser of the optimisation problem \eqref{eq202} if, and only if,
\begin{equation}\label{eq6}
\sum_{i = 1}^N g_i'(s^\star) = 0.
\end{equation}
\end{prp}

\begin{proof}
Our proof is based on the equivalence between \eqref{eq201} and \eqref{eq202}; that is, we exploit the fact that both problems are linked by a one-to-one change of variables. Indeed, it follows from this one-to-one mapping that $s^\star$ is the global minimiser of \eqref{eq202} if, and only if, the associated optimal travel times $t_i^\star = d_i/s^\star$ are the global minimisers of \eqref{eq201}. As the Lagrangian associated with \eqref{eq201} can be written as
\begin{equation}\label{eq204}
{\cal L}(t_i,\lambda_{ij}) = \sum_{i = 1}^N f_i(t_i) + \sum_{i>j} \lambda_{ij} \left( \frac{t_j}{d_j} - \frac{t_i}{d_i}\right)
\end{equation}
for some scalars $\lambda_{ij} \in \R$, $i > j$. From classic convex optimisation theory \cite{boyd_cvx}, the travel times $t_i^\star$ are optimal if, and only if, there exist Lagrange multipliers $\lambda_{ij}$ such that
\begin{equation}\label{eq205}
f_i'(t_i^\star) + \frac{1}{d_i}\left(\sum_{\ell < i} \lambda_{i\ell} - \sum_{\ell > i} \lambda_{\ell i}\right) = 0,
\end{equation}
together with the feasibility conditions $t_i^\star/d_i = t_j^\star/d_j$, $i > j$. Multiplying \eqref{eq205} by $d_i$, for each $i$, and summing all these equations up, it follows that the travel times $t_i^\star$ are optimal if, and only if,
\begin{equation}\label{eq206}
\sum_{i = 1}^N d_i f_i'(t_i^\star) = 0.
\end{equation}
The choice of optimal travel times $t_1,\cdots,t_N$ is unique, since each $f_i$ is strictly convex and so is their weighted sum, as $d_i > 0$ for $i = 1,\cdots,N$. Now, taking \eqref{eq6}, and the definition of $g$, we have that
\begin{equation}\label{eq207}
\sum_{i = 1}^N g_i'(s^\star) = -\frac{1}{(s^\star)^2} \sum_{i = 1}^N d_i f_i'\left(\frac{d_i}{s^\star}\right) = -\frac{1}{(s^\star)^2} \sum_{i = 1}^N d_i f_i'\left(t_i^\star\right),
\end{equation}
which implies that the proposed optimality condition in \eqref{eq6} is verified for some common speed $s^\star$ if, and only if, the associated travel times $t_i^\star = d_i/s^\star$ are optimal for \eqref{eq201}. Global optimality then follows from the definition of $g$. Uniqueness follows from the uniqueness of the optimal travel times. The proof is complete.
\end{proof}

\subsection{Optimisation and Algorithm}
Given the existence of this global solution, let us now focus on how to compute it in a distributed fashion. Following \cite{liu16}, we wish to solve this problem using an iterative procedure that converges to a recommended speed. To this end, we consider the iterative scheme
\begin{equation}\label{eq203}
\textbf{s}(k+1) = P(k) \textbf{s}(k) + G(\textbf{s}(k)) e,
\end{equation}
where $\{P(k)\}_{k \in \N} \subset R^{N \times N}$ is a sequence of row-stochastic matrices, $e \in \R^N$ is a vector with all entries equal to $1$, and $G \, : \, \R^N \to \R$ is a continuous function that verifies some assumptions, as we shall see in the sequel. Algorithms of this class have been extensively studied in the literature \cite{Korn09,Korn11,LiuAuto} and a contribution to the convex optimisation framework is given in \cite{liu16}. In this paper, we show that this algorithm is also applicable to solve a non-convex optimisation problem that presents some properties; one of these properties is the existence of a unique minimiser. \\

Note that there are two main components in the iteration \eqref{eq203}. The first component, which is the row-stochastic matrix $P(k)$, $k \in \N$, can be used to model the time-varying topology of the communication network among cyclists at every instance of time $k$. Roughly speaking, $P(k)$ is used to induce all components of $s$ to achieve a common value whereas the second one, given by a nonlinear function $G$, focuses on ensuring that some constraint must be verified. As in this paper our target is to achieve optimal consensus, $G$ must be chosen to achieve optimality at convergence. Our choice is, as in \cite{liu16},
\begin{equation}\label{eq209}
G(s) = -\mu \sum_{i = 1}^{N}g_i'(s_i),
\end{equation}
for all $s_i \in \R_+^\star$. For this particular choice, we may state the following theorem.

\begin{thm}
	Consider the optimisation problem \eqref{eq202}, the iteration \eqref{eq203} and the associated one-dimensional Lur'e system
	\begin{equation}\label{eq210}
	\begin{array}{rcl}
	y(k+1) & = & h(y(k)) , \\
	h(y) & \defeq & y + G(ye),
	\end{array}
	\end{equation}
	in which $G$ is the function defined in \eqref{eq209}, assumed to be continuous for $s_i > 0$, $i = 1,\cdots, N$. Suppose that $y^\star$ is a locally asymptotically stable fixed point of \eqref{eq210} and that $\big\{ P(k)\big\}_{k \in \N} \subset \R^{N \times N}$ is a strongly ergodic sequence of row-stochastic matrices. Then, $y^\star e$ is a locally asymptotically stable equilibrium point of \eqref{eq203}.
\end{thm}

For the proof of a similar result, see \cite{LiuAuto}. It is important to see that a point $y^\star$ is a fixed point for \eqref{eq210} if, and only if $h(y^\star) = y^\star$, which happens if, and only if $s^\star = y^\star e$ verifies \eqref{eq6}; that is, the fixed point of \eqref{eq203} is the optimal solution to \eqref{eq6}. Global convergence conditions shall be considered in future research. Hence, to construct the optimal solution to \eqref{eq6}, it only remains to choose the stochastic matrices $P(k)$, $k \in \N$, which are defined as
\begin{equation} \label{Pk}
P_{i,j}\left(k\right)=\left\{ \begin{array}{cc}
1-\sum_{j\in N_{k}^{i}}\eta_j, & \mbox{if }j=i,\\
\eta_j, & \mbox{if }j\in N_{k}^{i},\\
0, & \mbox{otherwise.}
\end{array}\right.,
\end{equation}
where $i,j$ are the entries' indexes of the matrix $P\left(k\right)$, and $\eta_j \in \mathbb{R}$ is a weighting factor. In this work, $\eta_j$ is chosen as $\frac{1}{\left| N_k^{i} \right| + 1}$ for simplicity, where $\left| \bullet \right|$ denotes cardinality, giving rise to an equal weight factor for all elements in $\textbf{s}\left(k\right)$.

\noindent \textbf{Comment: } Note that the positive scalar $\mu$ plays a key role in the (local) convergence of our algorithm and that it must be tuned by the designer. Global and local convergence conditions for this quasi-convex consensus setting shall be investigated in future research. \newline

\noindent \textbf{Comment: } Our main results are directed at consensus type applications where a basic type of ergodicity is assumed to hold. Clearly, this assumption is not always true. However, we note the following facts which are pertinent for applications in ITS, each of which make the assumption of strong ergodicity plausible.

\begin{itemize}
	\item[(i)] We are primarily motivated by ITS applications in which a group of bikes are travelling in close proximity to each other, thereby giving rise to connected communication graphs \cite{liu16}.
	\item[(ii)] Many applications of this type also operate a form of topology control to ensure either spatial or temporal connectivity. Details of one such algorithm is given in \cite{knorn2009framework}.
	\item[(iii)] If cyclists do not follow the suggested speed, and others do, then bikes will be come closer in space to each other, thereby making the graph more connected, and this will have the effect of making the graph strongly ergodic.
	\item[(iv)] Finally, the central agent can be used to send global information (other than derivatives), every so often, so as to make strong ergodicity even more likely.\newline
\end{itemize}

Now we propose the following optimal distributed consensus algorithm for solving optimisation problem \eqref{eq202} as follows. Note that the proposed algorithm can be implemented in a privacy preserving manner. Interest reader is referred to \cite{liu16} for more details.


\begin{algorithm}[htbp]
	\caption{Optimal Distributed Consensus Algorithm}
	\begin{algorithmic}[1]
		\For{$k=1,2,3,..$}
		\For{$i \in \left\lbrace 1,2,\ldots, N \right\rbrace$}
		\State Get $G(\textbf{s}(k))$ from the base station.
		\State Get $s_j\left(k\right)$ from all neighbours of e-bike $i$.
		\State Do $q_i\left(k\right)=\eta_i\cdot \sum\limits_{j\in N_k^{i}}\left(s_j\left(k\right)-s_i\left(k\right)\right)$.
		\State Do $s_i\left(k+1\right) = s_i\left(k\right) + q_i\left(k\right) - \mu \cdot \tilde{F}\left(k\right)$.
		\EndFor
		\EndFor
	\end{algorithmic}
	\label{alg:Algorithm1}
\end{algorithm}

\section{Simulation} \label{Simulation}

In this section, we present some preliminary results obtained from Matlab simulations. In particular, we evaluate the performance of our proposed algorithm in two scenarios at different cities where the background pollutant level is significantly different for cyclists. In both scenarios, we assume that there are 15 bikes sharing a common route.

\subsection{Construction of Risk Functions}

\begin{figure}[htbp]
	\begin{center}
		\includegraphics[width=3.4in, height = 2.8in]{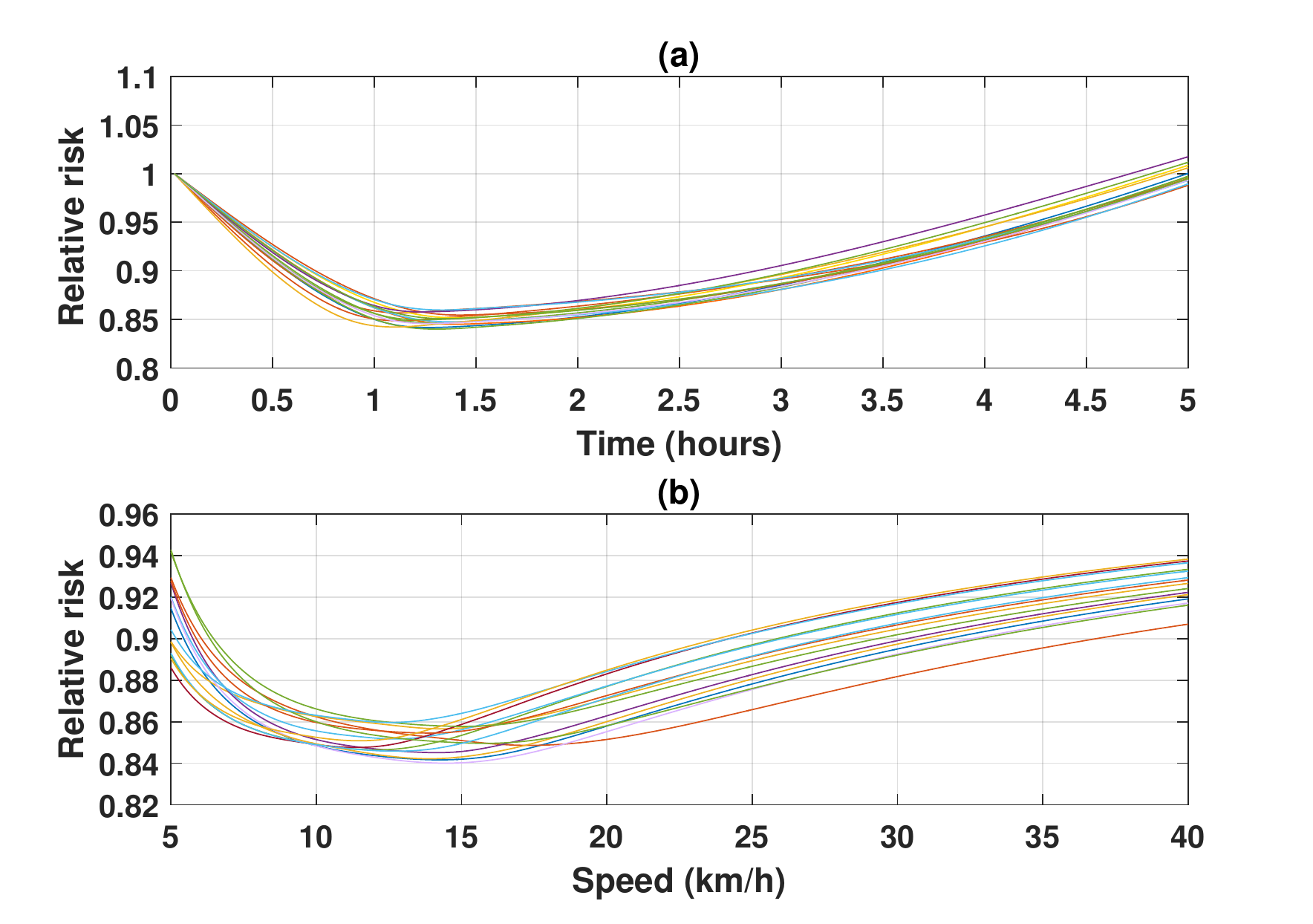}
		\caption{Risk functions $f_i$ (a) and $g_i$ (b) for 15 cyclists in low polluted city cycling scenario.}\label{CFGood}
	\end{center}
\end{figure}

\begin{figure}[htbp]
	\begin{center}
		\includegraphics[width=3.4in, height = 2.8in]{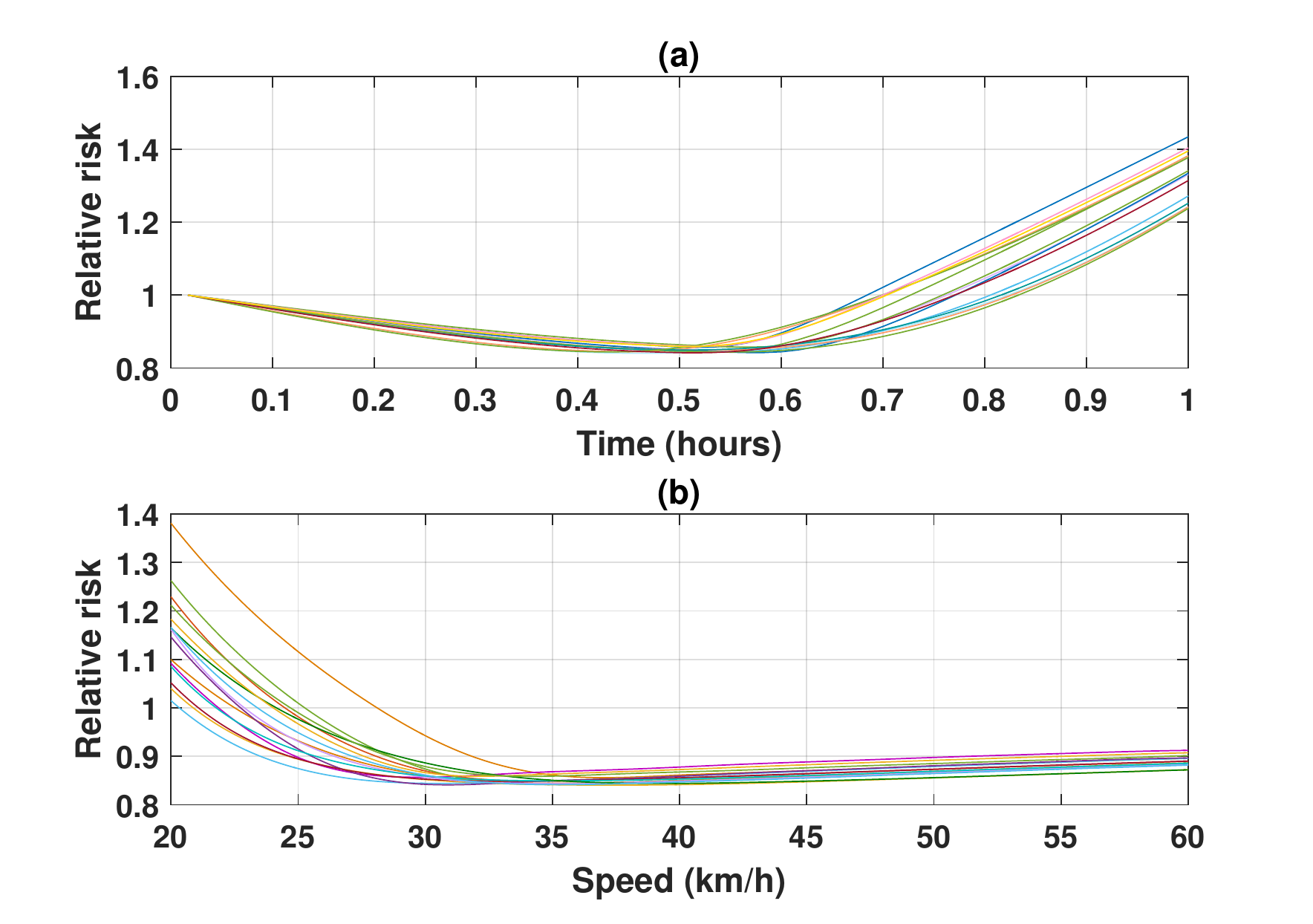}
		\caption{Risk functions $f_i$ (a) and $g_i$ (b) for 15 cyclists in high polluted city cycling scenario.}\label{CFIndia}
	\end{center}
\end{figure}

In this section, we construct risk functions for a group of cyclists in different city cycling scenarios. In particular, we consider a low polluted city cycling scenario with PM 2.5 background level set to $50 ug/m^3$, and a high polluted city cycling scenario with PM 2.5 background level set to $153 ug/m^3$ \cite{tainio}. In both cases, the basic profile of each risk function is chosen according to the curve shown in Figure \ref{risk}. However, considering the fact that cyclists may have different level of fitness and may also have access to e-bikes, we factor in this randomness by adjusting both tipping point and breakeven point of each cyclist's risk function in a small area. As a result, the risk functions $f_i(t_i)$ of all cyclists are presented in the subplot (a) of Figure \ref{CFGood} for the low polluted scenario, and subplot (a) of Figure \ref{CFIndia} for the high polluted scenario. Note that, each function $f_i$ is fitted as a strictly convex function by using spline interpolation \cite{de2005spline}. \\

In order to model the risk function $g_i(s_i)$ for each cyclist, we assume that the daily travelling distances of all cyclists are distributed uniformly between 15 and 20 km in both scenarios. Thus, by changing of variables, we can easily obtain the corresponding risk function $g_i(s_i)$ given both $f_i(t_i)$ and $d_i$. Each function $g_i$ is then fitted again using spline interpolation, and the fitted curves are presented for low and high polluted city scenarios in subplot (b) of Figure \ref{CFGood} and \ref{CFIndia}, respectively. Now it can be seen from both figures that the functions turn out to be strictly quasi-convex functions instead of strictly convex functions.

\subsection{Algorithm Evaluation}

\begin{figure}[htbp]
	\begin{center}
		\includegraphics[width=3.4in, height = 2.8in]{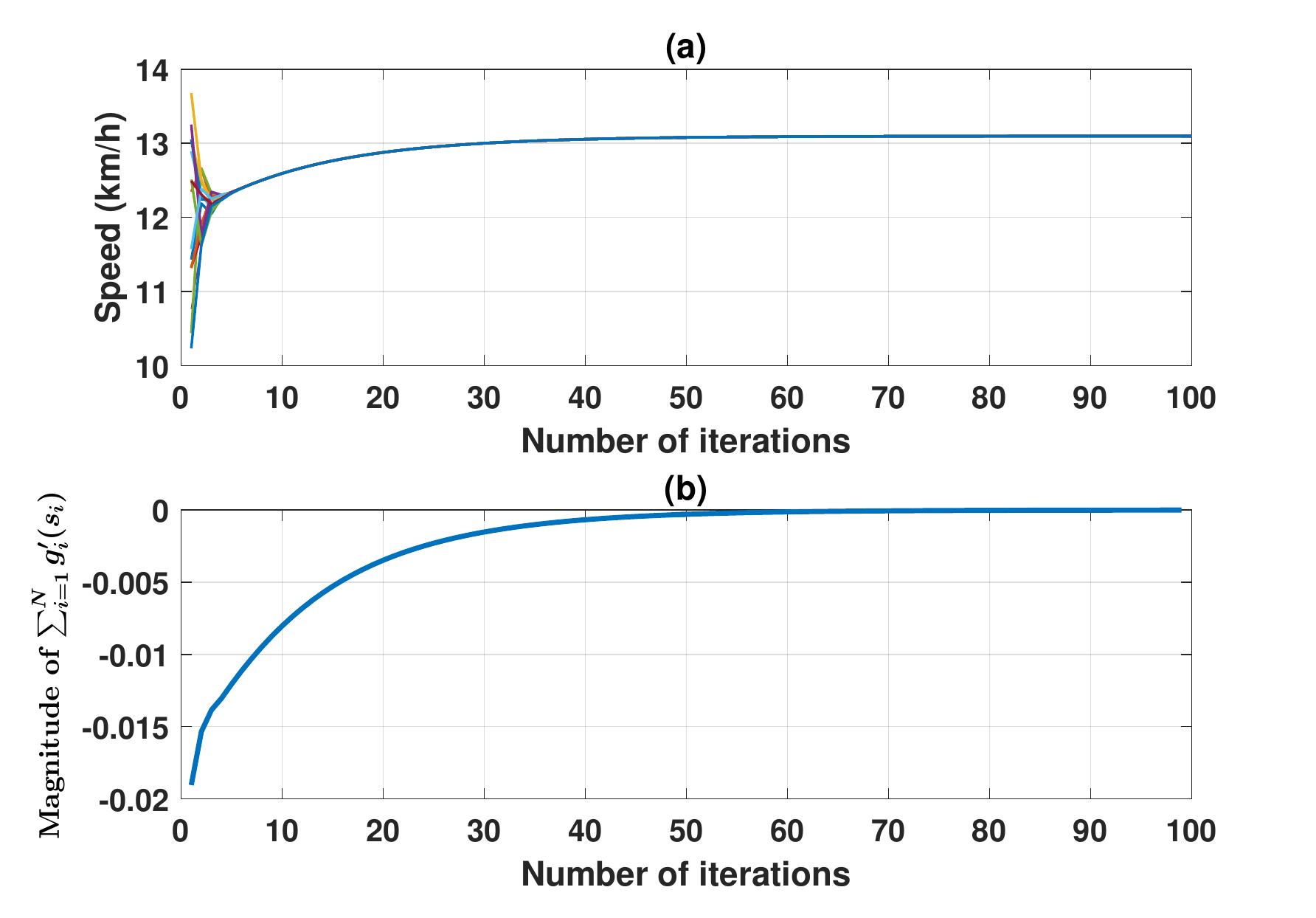}
		\caption{Recommended speeds of cyclists converge to consensus (a) and optimum (b) in the low polluted city cycling scenario.}\label{SpeedConvergeGood}
	\end{center}
\end{figure}

\begin{figure}[htbp]
	\begin{center}
		\includegraphics[width=3.4in, height = 2.8in]{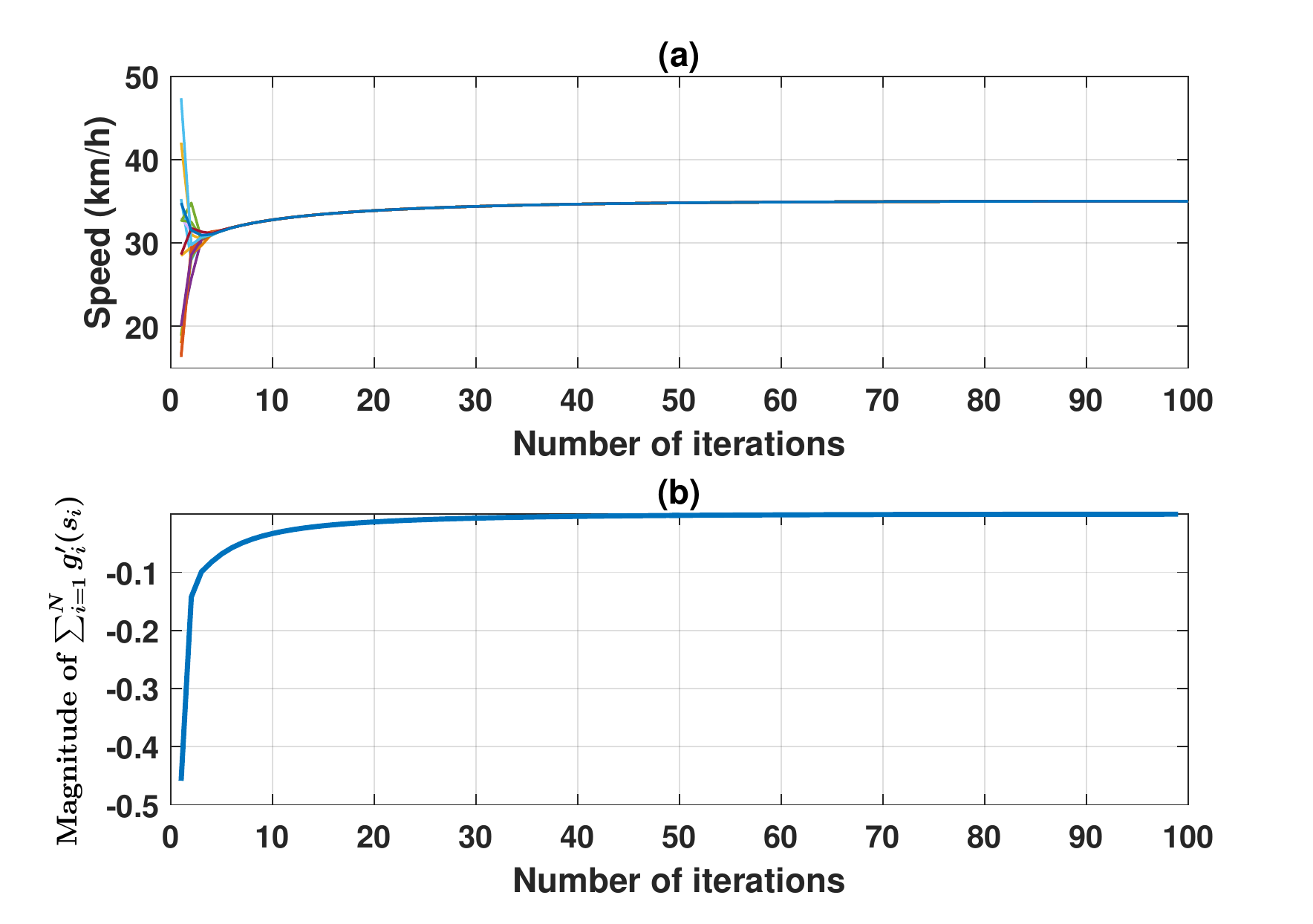}
		\caption{Recommended speeds of cyclists converge to consensus (a) and optimum (b) in high polluted city cycling scenario.}\label{SpeedConvergeIndia}
	\end{center}
\end{figure}

In this section, we evaluate the performance of the proposed optimal distributed consensus algorithm in two scenarios with respect to different background pollutant levels. The simulation results for both low and high polluted city cycling scenarios are illustrated in Figure \ref{SpeedConvergeGood} and Figure \ref{SpeedConvergeIndia} respectively. In the case of low polluted cycling, the initial speeds of all cyclists are assumed to be uniformly distributed between 10 and 15 km/h as shown in the subplot (a) of Figure \ref{SpeedConvergeGood}. After running less than 50 iterations, the recommended speeds successfully converge to 13.3 km/h for all cyclists, and the subplot (b) of Figure \ref{SpeedConvergeGood} further validates the optimality of the algorithm. \\

Similarly, it can be seen from Figure \ref{SpeedConvergeIndia} that the proposed algorithm also helps recommended speeds converge to consensus very efficiently in the high polluted cycling scenario. However, the optimal speed for all cyclists in this scenario becomes around 35 km/h, which is nearly three times than the optimal speed calculated in the low polluted scenario. Therefore, it makes more sense for cyclists in high polluted city to maintain their high speeds when cycling by using more power assistance.

\section{Conclusions} \label{Conclusion}

In this paper, we have presented a novel design of the speed advisory system for group of cyclists. The system is operated by sending advisory speeds to bikes in order to minimise the overall health risks of the group during cycling. The underlying mechanism of the system, which extends our previous theoretical findings, is to achieve consensus and optimality on the recommended speeds in an iterative manner which potentially preserves privacy of cyclists. Our proposed approach paves the way for cyclists to effectively use their electric bikes when travelling as a group while substantially improving their cycling experience without imposing additional health damage. We have implemented the proposed system in cities with both clean and dirty background pollutants. Our results show that cyclists living in less clean air environment would generally require more power assistance than the other since they have faster cycling requirement to keep better health conditions.

\section*{Acknowledgment}
The authors gratefully acknowledge funding for this research provided by Science Foundation Ireland under grant 11/PI/1177. The work of Matheus Souza was supported by Funda\c{c}\~ao de Amparo \`a Pesquisa do Estado de S\~ao Paulo (FAPESP) grant 2016/19504-7.

\bibliographystyle{IEEEtran}
\bibliography{refs}             

\end{document}